\documentstyle[aps]{revtex}
\begin{document}
\draft
\title{On the conical refraction of hydromagnetic waves in plasma
with anisotropic thermal pressure}
\author{David Tsiklauri \footnote{On leave from the Department of Theoretical
Astrophysics, Abastumani  Astrophysical
Observatory, Kazbegi str. $N.~2^{a}$, Tbilisi 380060, Republic of Georgia}}
\address{
Physics Department, University of
Cape Town, Rondebosch 7700, Republic of South Africa}
\date{\today}
\maketitle
\begin{abstract}
A phenomenon analogous to the conical refraction widely known in the
crystalooptics and crystaloacoustics is discovered for the
magnetohydrodynamical waves in
the collisionless plasma with anisotropic thermal pressure.
Angle of the conical refraction is calculated for the
medium under study which is predicted to be $18^{\circ}26^{\prime}$.
Possible experimental corroborating of the discovered phenomenon
is discussed.
\end{abstract}

\pacs{52.35.Bj, 42.15.Gs}

\section{Introduction}
A phenomenon of the conical refraction --- a peculiar kind of light
refraction in biaxial crystals taking place when a direction of a light
beam coincides with some of the optical axes (binormals) of the
crystal, was predicted theoretically by W.R. Hamilton even in 1832.
He discovered this phenomenon basing on the Huygens--Fresnel principle
(formulated 17 years earlier) applied for the biaxial crystals. In 1833
H. Lloyd proved experimentally existence of the phenomenon. Observation
of the inner conical refraction in biaxial crystal is possible if the
parallel, narrow beam of natural (non-polarized) light falls
along the optical axis (binormal) on a plate cut out from the crystal
perpendicularly to the binormal. Propagating through the crystal
the beam will form an empty cone with permanently changing linear
polarization. Further, the phenomenon of the conical refraction
was discovered for the elastic waves propagating along the acoustical
axis of the crystal. For example, the inner conical refraction may
take place when purely shearing waves propagate through the crystal
along the third order symmetry axis (principal axis of symmetry
[001] in triagonal crystals; direction [111] along the diagonal of
a cube in the cubical crystals). In certain cases phenomenon of
the acoustical conical refraction may be rather pronounced.
The angle of the conical refraction (i.e., angle of deflection of
the beam from the acoustical axis of the crystal) may have
significant magnitude, for instance, in calcit
$CaCO_3{\approx}30^{\circ}$, in quartz ${\approx}17^{\circ}$ etc.

Resuming aforesaid, the phenomenon of the conical refraction in the
crystals (optical and acoustical) is known for quite a long time and
hence is completely studied. However, further display of the
phenomenon is rather surprising: in Ref. [1] there was theoretically
discovered the phenomenon analogous to the inner conical refraction for
{\it magnetosonic} waves propagating in the isotropic plasma which is
considered as a perfectly conducting fluid with adiabatical
perturbations in it (see for brief description of the effect
more accessible Ref. [2]).
In particular [1], if the Alfven velocity
coincides with the speed of sound then in the case of propagating
of a quasiplane magnetosonic wave along the uniform magnetic field
the beams corresponding to the wave will form an empty cone with the
opening angle $53^{\circ}08^{\prime}$, i. e. the angle of conical refraction
for the magnetosonic waves in isotropic plasma was predicted to be
$26^{\circ}34^{\prime}$.

In the present paper, while considering adiabatical
disturbances in collisionless plasma contained in a uniform external
magnetic field and hence described by the theory of Chew, Goldberger
and Low [3,4,5], is discovered phenomenon analogous to the inner conical
refraction. Calculations showed that pending angle of the conical
refraction for the magnetohydrodynamical (MHD) waves in the plasma with
anisotropic thermal pressure should be $18^{\circ}26^{\prime}$.
The aim of this paper is to show {\it theoretically} that the phemomenon
analogous to the inner conical refraction should also exist in anisotropic
plasma medium and to encourage other authors to prove {\it experimentally}
existence of the phenomenon.  The beams are interpreted as lines along which
the group velocity
vector is directed.

In the concluding part of the paper there are discussed possibilities of
experimental corroborating of the discovered phenomenon.

\section{Main consideration}

It is known [4,5] that the linear MHD waves existing in the plasma
described by the anisotropic thermal pressure obey to the following
dispersion relation:
$$
\omega^2={{k^2}\over{2\rho_0}}
\Biggl[{{B_0^2}\over{4\pi}}+p_{\perp}(1+\sin^2{\theta})+2p_{\parallel}
\cos^2{\theta}
$$
$$
{\pm}{\sqrt{{\left({{{B_0^2}\over{4\pi}}+p_{\perp}(1
+\sin^2{\theta})-
4p_{\parallel}\cos^2{\theta}}\right)}^2+
4p_{\perp}^2\sin^2{\theta}\cos^2{\theta}}}\Biggr] \eqno (1)
$$
We further can rewrite eq. (1) in more convenient form introducing
notations similar to those ones from Ref. [1]:
$$
a={{B_0}\over{\sqrt{4\pi\rho_0}}}, \,\,\,
c_\perp=\sqrt{{p_\perp}\over{\rho_0}}, \,\,\,
c_\parallel=\sqrt{{p_\parallel}\over{\rho_0}}
$$
where $a$ is the Alfven
velocity, $c_{\perp}$ and $c_{\parallel}$ are the sound speeds measured
in perpendicular and parallel directions in respect to the uniform,
external magnetic field ${\vec B}_0$. Doing so, we get:
$$
V^2={{{1}\over{2}}{\left[{a^2+c_{\perp}^2(1+\sin^2{\theta})+2c_{\parallel}^2
\cos^2{\theta}{\pm}{\sqrt{{\left({a^2+c_{\perp}^2(1+\sin^2{\theta})-
4c_{\parallel}^2\cos^2{\theta}}\right)}^2+
4c_{\perp}^4\sin^2{\theta}\cos^2{\theta}}}}\right]}} \eqno (2)
$$
where $V=\omega/k$ denotes phase velocity of the waves.  Traditionally, here
$\theta$
denotes an angle between ${\vec B}_0$ and wave vector ${\vec k}$.

Let us consider the special subcase of eq. (2) assuming that condition
$c_{\perp}={\sqrt{2}c_{\parallel}}$ is fulfilled in the medium under
study. It is worthwhile to note that the latter assumption leads to
existence of so called "angular point" [1] on
the surfaces formed by the phase velocities (both fast and slow MHD
waves) just in the point of intersection of the ones with the
direction of the magnetic field  ${\vec B}_0$. Such a location
of angular point, in turn, is responsible for the existence of the
effect of conical refraction in the medim under study. The main guess was
to find such a condition which would led to presence of the conical
refraction in the {\it anisotropic} plasma medium. i.e. to discontinuous
behavour of the vector of group velocity when $\theta$ tends to zero
(see below). Since possessing such a condition we can rewrite eq.(2)
as a following:
$$
V^2={{{1}\over{2}}{\left[{a^2+2c_{\perp}^2{\pm}
{\sqrt{{\left({a^2+2c_{\perp}^2}\right)}^2-
4{\left({\left[{{{3}\over{2}}(a^2+2c_{\perp}^2)-c_{\perp}^2}\right]
c_{\perp}^2\cos^2{\theta}-
{{5}\over{4}}c_{\perp}^4\cos^4{\theta}}\right)}}}}\right]}} \eqno (3)
$$
Basing on the latter expression we may calculate the group velocity
of the waves. In general, the group velocity ${\vec u}$ has physical
meaning of the wave packet propagation speed and can be found by simple
formula ${\vec u}={\partial}{\omega}/{\partial}{\vec k}$. The derivation
of (3) results:
$$
{\vec u}=
{V{{{\left[{{\left({1-{\left[{(5c_{\perp}^4)/(4V^4)}\right]}
\cos^4{\theta}}\right)}}\right]}{\vec N}-
{\left[{{\left({{\cos{\theta}}/{V^4}}\right)}
{\left({{\left[{{{3}\over{2}}(a^2+2c_{\perp}^2)-c_{\perp}^2}\right]
c_{\perp}^2-
{{5}\over{2}}c_{\perp}^4\cos^2{\theta}}}\right)}}\right]}{\vec A}}\over{1-
{\left[{{\left[{{{3}\over{2}}(a^2+2c_{\perp}^2)-c_{\perp}^2}\right]
c_{\perp}^2\cos^2{\theta}-
{{5}\over{4}}c_{\perp}^4\cos^4{\theta}}}\right]}/V^4}}} \eqno (4)
$$
where we introduced unit vectors ${\vec N}={\vec k}/|{\vec k}|$
and ${\vec A}={\vec B}_0/|{\vec B}_0|$. To realize the physical picture
geometrically it is worthwhile to state here following reasonings [1]:
let multiply scalarly the latter
formula by ${\vec N}$:
$$
({\vec u}{\vec N})=V \eqno(5a)
$$
i.e. projection of group velocity on the direction of wave nornal is equal to
the phase velocity in the same direction. Multiplying eq.(5a) by $k$ one can
get:
$({\vec u}{\vec k})=\omega$. Hence, for the infinitesimal changes of ${\vec
u}$,
${\vec k}$ and $\omega$ the following relation is valid:
$({\vec u}\delta{\vec k})+({\vec k}\delta{\vec u})=\delta \omega$.
According to the definition of group velocity
$({\vec u}\delta{\vec k})=\delta \omega$. Therefore, $({\vec k}\delta{\vec
u})=0$,
or in other terms
$$
({\vec N}\delta{\vec u})=0 \eqno(5b)
$$
The vector $\delta{\vec u}$
is located in the plane tangential to the surface (so called
beam surface [1]) formed by the ends of vectors of the group velocities drawn
away from the arbitrary point $O$. Therefore, eq.(5b) implies that the plane,
tangential to the beam surface, is
perpendicular to the corresponding wave nornal. i.e. the beam surface is
the envelope for the plane wave fronts, which were propagated from reference
point $O$ in every directions per unit time. The latter statement in
combination with
eq.(5a) completely determines geometrical relationship between surface of
phase velocities with the beam surface. However, it should be stressed that
this geometrical relationship is valid only for the waves which phase
velocities depend only on the direction and not on the length of the
wave vector ${\vec k}$. It is enough to show that in such case eq. (5a) is
valid.
Let $V$ (phase velocity) be arbitrary function of wave vector ${\vec k}$.
Then $\omega=kV({\vec k})$ and hence,
${\vec u}=\partial \omega/ \partial {\vec k}=V({\vec k}){\vec k}/k+
k \partial V/ \partial {\vec k}$. It is necessary and sufficient for
condition (5a) to be hold that the $V$ must satisfy the following
requirment: ${\vec k} \partial V/ \partial {\vec k}=0$ or that is the same
as $k_i \partial V/ \partial k_i=0$. Therefore, one can conclude that $V$ must
be homogeneous function of $k_i$ components of the zeroth order.
The latter statement is equivalent to the requirement that function $V$
depends only on the direction of wave vector ${\vec k}$ and not on its
length. As it can be seen clearly the dirpersion relations (1) and of course
(2) and (3) satisfy this condition.

It is known [1] that since possessing surface formed by phase velocities
one can construct geometrically the beam surface (formed by the ends of
vectors of the group velocities) and {\it vice versa}. But instead of
doing this it more convenient to carry out analitical calculation.
Let us describe a direction of the group velocity ${\vec u}$ by an
angle $\varphi$ between ${\vec u}$ and ${\vec B}_0$ (i.e., between ${\vec u}$
and ${\vec A}$).  One can determine the angle
$\varphi$ by the formula:
$$
\tan \varphi = {{{\vec u} \times {\vec A}}\over{{\vec u} \cdot {\vec A}}}
$$
which with the aid of eq. (4) yields:
$$
\tan{\varphi}={{{{\left[{{\left({1-{\left[{(5c_{\perp}^4)/(4V^4)}\right]}
\cos^4{\theta}}\right)}}\right]}}\over{{1-
{\left[{{\left[{{{3}\over{2}}(a^2+2c_{\perp}^2)-c_{\perp}^2}\right]
c_{\perp}^2-
{{5}\over{2}}c_{\perp}^2\cos^2\theta}+
{{5}\over{4}}c_{\perp}^4\cos^4{\theta}}\right]}/V^4}}}\tan{\theta}} \eqno(6)
$$
By means of the eq. (6) we completely determined the direction
of the group velocity ${\vec u}$ whereas its module can be calculated
by expression:
$$
{|{\vec u}|} ={V/\cos({\varphi}-{\theta})} \eqno (7)
$$
It is worthwhile to note that the eq. (7) can be obtained by using
self-evident connection (5a) between the group velocity ${\vec u}$ and phase
velocity $V$.

It is simple to show that expressions (4) and (6) have uncertain
behavior as $0/0$ when $a=c_{\perp}$ (i.e., when
$a=c_{\perp}=\sqrt{2}c_{\parallel}$) and $\theta$ tends to zero.
Removing the uncertainty yields:
$$
\tan\varphi(+0)={\lim_{\theta\to0}{\tan\varphi(\theta)}}=
{{\pm}{{1}\over{3}}} \eqno(8)
$$
$$
u(+0)={{{\sqrt{10}}\over{3}}c_{\perp}} \eqno(9)
$$
in eq. (8) sign "+" and "-" correspond to the slow and fast MHD waves
respectively. The above treatment shows that when the wave normal
${\vec N}$ tends to reach a direction parallel to ${\vec B}_0$
(i. e. when $\theta$ tends to zero), vector of group velocity ${\vec u}$
does not takes the same direction, but its limiting position
(when $\theta\to0$) is the
surface of an empty cone with opening angle
$2{\varphi}(+0)=2\arctan(1/3)=36^{\circ}52^{\prime}$. When $\theta$ is zero
exactly then
${\vec u}$ is strictly parallel to ${\vec B}_0$. Such a discontinuous
behavior of vector ${\vec u}$ can be explained by means of existence of
so called [1] angular point which in the case of the anisotropic plasma
exists when $a=c_{\perp}=\sqrt{2}c_{\parallel}$. (Note that for the
waves in isotropic plasma the latter condition was rather simple [1]
$a=c$ where $c$ denotes ordinary sound speed). So, when
$a=c_{\perp}=\sqrt{2}c_{\parallel}$
the angular point on the surfaces formed by the phase velocities
(both fast and slow MHD waves) is located in the point of intersection
of the ones with the direction of the magnetic field  ${\vec B}_0$.
Which, in turn leads to the discontinuous behaviur of ${\vec u}$ and
causes the hole effect.
In other terms the angular point is nothing
else then point of intersection of the phase velocities of the fast and slow
MHD waves. One can see from the Fig. 1 that when the condition
$a=c_\perp$ is not satisfied than the polars of phase velocities
(according to the dispersion relation (3)) do not
have any points of intersection.
It  follows from the eq. (2) that when $a=c_{\perp}=\sqrt{2}c_{\parallel}$,
then
$$
V^2={{{3a^2}\over{2}}{\left[{1 \pm {{\sin \theta}\over{3}} \sqrt{4+
5\sin^2 \theta}}\right]}}. \eqno (10)
$$
{}From the eq.(10) we gather that when $\theta=0$ then $V_+=V_-$ and hence
the phase polars have two points of intersection which can be clearly seen
from the Fig.2. It is
apparent that in these points derivative of the phase
velocities has uncertain behaviour. Removing of this uncertainity was done
above. Result is clear now: in the anisotropic plasma for the linear MHD waves
there should exist the effect analogous to the conical refraction and we
worked out detailed explanation of the phenomenon.

\section{Discussion}

Here in this section we discuss possibilities of experimental
testing of existence of the phenomenon. Strictly speaking, the plane
wave is idealization and real physical significance has quasiplane wave
with small but finite angle of aperture. It should be emphasized
that no real physical object corresponds to the wave with wave normal
${\vec N}$ exactly parallel to ${\vec B}_0$ ($\theta=0$), whereas in
the case of propagation of quasiplane MHD wave with small but finite
angle of aperture ($\theta\to0$) the narrow beam of such
quasiplane waves, as it was shown above, will form an empty cone with opening
angle
$2{\varphi}(+0)=36^{\circ}52^{\prime}$ if
$a=c_{\perp}=\sqrt{2}c_{\parallel}$ condition is fulfilled.
We may suppose the following experiment for demonstration
of the phenomenon analogous to the conical refraction: Let us consider
two collisionless plasma  media contained in a uniform external
magnetic field ${\vec B}_0$. The media are separated by means of
the plane perpendicular to the ${\vec B}_0$. Let choose external conditions in
the second
medium so that to satisfy the condition
$a=c_{\perp}=\sqrt{2}c_{\parallel}$ permanently. If we place a diaphragm
between
the first and second media then the beam of MHD waves propagating
from the first medium towards the second one, just after passing
the diaphragm will deflect from the direction of ${\vec B}_0$ and form
an empty cone with angle of opening
$2{\varphi}(+0)=36^{\circ}52^{\prime}$.

It is worthwhile to note that the effect of the conical refraction
may be used for plasma diagnostics. It is clear that the due to the effect
of the conical refraction intensity of wave which is registered by a detector
placed on the path of propagation of the wave will show steep drop
for certain value of the magnetic field which satisfies the condition
$$
{{B_0^2}\over{{4 \pi }}}= p_{\perp}=2 p_{\parallel}
$$
Such a technique was proposed for isotropic plasma case in Ref.[6]

For the author is unknown whether the phenomenon analogous
to the inner conical refraction of MHD waves in the anisotropic plasma was
observed experimentally yet. However, one should take into account the
difficulties
in prepation of a plasma medium with permanently satified condition
$a=c_{\perp}=\sqrt{2}c_{\parallel}$.
The aim of this paper is to stimulate other authors, in spite of the
techinacal difficulties, to prove the existence of the effect experimentally.

\section{Acknowledgements}
This research was supported, in part, by  South African Foundation (FRD)
for Research Development. It is my pleasure to express my gratitude to
Prof. R.D. Viollier and Theoretical Physics Group of the Physics
Department of the University of Cape Town for warm hospitality and
creative atmosphere.

\newpage

\setlength{\unitlength}{0.240900pt}
\ifx\plotpoint\undefined\newsavebox{\plotpoint}\fi
\sbox{\plotpoint}{\rule[-0.200pt]{0.400pt}{0.400pt}}%

\vskip 2cm
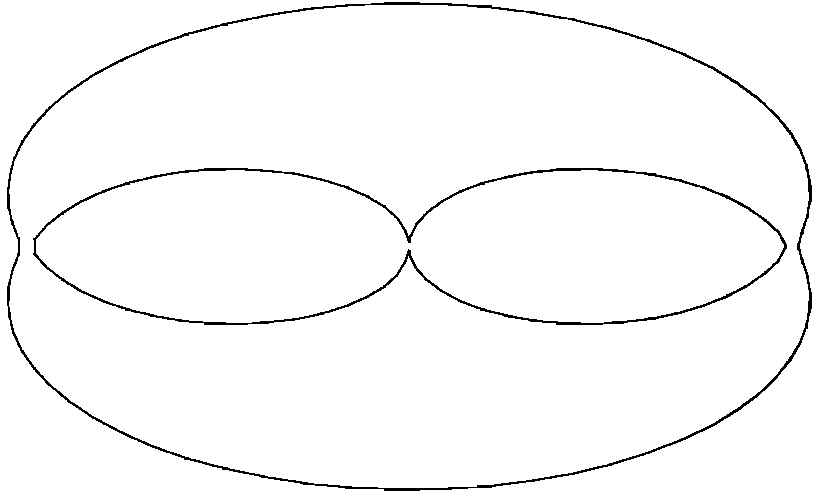
Fig. 1, graph of phase polars for slow and fast MHD waves
according to eq. (3) for the case when ${a/c_\perp} =0.9$
i.e. near the degeneracy.
\newpage


\setlength{\unitlength}{0.240900pt}
\ifx\plotpoint\undefined\newsavebox{\plotpoint}\fi
\sbox{\plotpoint}{\rule[-0.200pt]{0.400pt}{0.400pt}}%
%
\vskip 2cm

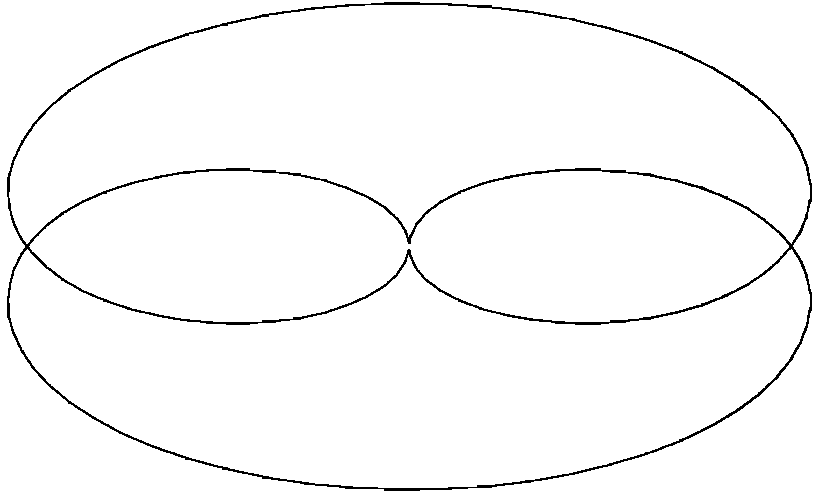
Fig. 2, graph of degenerate phase polars for slow and fast MHD waves
according to eq. (3) or as well as eq.(10) for the case when ${a/c_\perp} =1$

\begin{references}
\item
D.V. Sivukhin, Magnitnaia gidrodinamica, {\bf 1}, 35, 1966. (in
Russian)
\item
A. I. Akhizer and Akhiezer, {\it Plasma Electrodynamics}, vol. 1:
{\it Linear Theory}, (Pergamon Press, London, 1973)
\item
G. Chew, M. Goldberger, and F. Low, Proc. R. Soc. London ser.
     {\bf A 236}, 112 (1956).
\item
N. A. Krall and A. W. Trivelpiece, {\it Principles of Plasma
     Physics} (McGraw-Hill, New-York, 1973).
\item
F. F. Cap, {\it Handbook of Plasma Instabilities}, vol. 1,
(Academic Press, New York, 1976), p. 213
\item
R.V. Polovin and K.P. Cherkasova, {\it High-frequensy Behaviour of Plasma}
(Naukova Dumka, Kiev, 1967) (in Russian)
\end{references}
\end{document}